\documentclass{jnmp01b}

\usepackage{amsmath}

\numberwithin{equation}{section}

\allowdisplaybreaks

\newcounter{rostnr}
\newenvironment{roster}%
{\begin{list}{\textup{(\arabic{rostnr})}}%
    {\labelwidth=1.5em \itemsep=0pt \leftmargin=3em \usecounter{rostnr}}}%
{\end{list}}

\newtheorem{theorem}{Theorem}

\theoremstyle{definition}
\newtheorem{example}{Example}
\newtheorem*{remark}{Remark}
\newtheorem*{comments}{Comments}

\makeatletter
\DeclareRobustCommand{\primfrac}[1]{%
  \PackageWarning{amsmath}{%
Foreign command \@backslashchar#1; %
\protect\frac\space or \protect\genfrac\space should be used instead%
  }
  \global\@xp\let\csname#1\@xp\endcsname\csname @@#1\endcsname
  \csname#1\endcsname
}
\makeatother

\begin{document}

\renewcommand{\evenhead}{I.Z.\ Golubchik and V.V.\ Sokolov}
\renewcommand{\oddhead}{Operator Yang-Baxter Equations
and Nonassociative Algebras}


\thispagestyle{empty}

\begin{flushleft}
\footnotesize \sf
Journal of Nonlinear Mathematical Physics \qquad 2000, V.7, N~2,
\pageref{firstpage}--\pageref{lastpage}.
\hfill {\sc Article}
\end{flushleft}

\vspace{-5mm}

\copyrightnote{2000}{I.Z.\ Golubchik and V.V.\ Sokolov}

\Name{Generalized Operator Yang-Baxter Equations, Integrable 
ODEs and Nonassociative Algebras}

\label{firstpage}

\Author{I.Z.\ GOLUBCHIK~$^\dag$ and V.V.\ SOKOLOV~$^\ddag$}

\Adress{$^\dag$ Bashkirian Pedagogic Institute,  
  October Revolution st.\ 3 a, Ufa 450000, Russia \\[2mm]
  $^\ddag$ Center for Nonlinear Studies, 
  at Landau Institute for Theoretical Physics, \\
~~Russian Academy of Sciences,
  Kosygina st.\ 2, Moscow, 117940, Russia\\
~~Email: sokolov@landau.ac.ru
}

\Date{Received December 27, 1999; Revised March 9, 2000; 
Accepted March 10, 2000}

\begin{abstract}
\noindent
Reductions for systems of ODEs integrable via the standard factorization 
method (the Adler-Kostant-Symes scheme) or the generalized factorization 
method, developed by the authors earlier, are considered. Relationships 
between such reductions, operator Yang-Baxter equations, and some kinds of 
non-associative algebras are established.
\end{abstract}


\section{Introduction}

The factorization method (see [1]) is used to integrate 
a system of ODEs of the form
\begin{equation}
  q_t= [q_{+} \ ,q], \qquad q(0)=q_0.\label{0.1}
\end{equation}
Here $q(t)$ belongs to a Lie algebra $\frak G$, which is decomposed 
(as a vector space) into a direct sum 
\begin{equation}
  \frak G = \frak G_+ \oplus \frak G_- \label{0.2}
\end{equation}
of its subalgebras $\frak G_+$ and $\frak G_-$. The projection of $q$ 
onto $\frak G_+$ parallel to $\frak G_-$ is denoted by $q_+$. 
For simplicity, we assume that $\frak G$ is embedded into a matrix algebra.

The solution of (\ref{0.1}) can be written in the form 
\begin{equation}
  q(t)=A(t) q_0 A^{-1}(t).\label{0.3}
\end{equation}
The matrix-valued function $A(t)$ in (\ref{0.3}) is defined from 
the following factorization problem
\begin{equation}
  A^{-1}B = \exp (-q_0 t), \qquad A \in G_+, \qquad B \in G_-,
  \label{0.4}
\end{equation}
where $G_+$ and $G_-$ are the Lie groups of $\frak G_+$ and $\frak G_-$.
It is well known that for small $t$ the functions $A(t)$ and $B(t)$ are 
uniquely determined.
If the groups $G_+$ and $G_-$ are algebraic then the conditions
\begin{equation*}
  A \in G_+, \qquad A \exp (-q_0 t) \in G_-
\end{equation*}
are equivalent to a system of algebraic equations for the entries of $A(t)$.  

It follows from (\ref{0.3}) that if initial data $q_0$ for (\ref{0.1}) belongs to
a $\frak G_+$-module $M$, then $q(t) \in M$ for any $t$. Such a specialization
of (\ref{0.1}) gives rise to a system which we call $M$-{\bf reduction}. 

Let us denote by $R: M\rightarrow \frak G_+$ the projector onto     
$\frak G_+$ parallel to $\frak G_-$. In terms of $R$ the $M$-reduction 
can be written as follows
\begin{equation}
  m_t=[R(m), \, m], \qquad m\in M. \label{0.5}
\end{equation}
As we demonstrate in Section 2, in important cases related to 
$\Bbb Z_2$-graded Lie algebras, the operator 
$R$ satisfies one of the generalized operator Yang-Baxter equations 
(see [2]-[4]). Vice versa, if $R: \frak A \rightarrow \frak A$ satisfies 
one of several kinds of operator Yang-Baxter equation on a Lie algebra 
$\frak A$, then equation (\ref{0.5}) can be regarded as an $\frak A$-reduction
of  (\ref{0.1}) for some $\frak G \supseteqq \frak A.$

The component form of (\ref{0.5}) is a system of quadratic ODEs.
There are a lot of works where such systems are studied by the methods 
based on the analysis of singularities of solutions (see [5],[6]). 
We believe that the concrete systems that appear in Section 2 as examples 
could be used for examining some of the conjectures in the literature 
about singularity analysis. It would be interesting also to investigate 
which types of singularities are possible for systems integrable by the 
generalized factorization method [7].

In Sections 3 and 4 we investigate properties of an algebraic operation $*$ 
on $M$, defined by 
\begin{equation}
  m*n=[R(m), \, n]. \label{0.6}
\end{equation}
In terms of this operation the $M$-reduction (\ref{0.5}) takes the form    
\begin{equation}
  m_t=m*m. \label{0.7}
\end{equation}
It turns out that the operation 
$*$ can be described by algebraic identities for several
examples of interest. Some generalizations of 
non-associative algebras such as Lie, left-symmetric and Jordan algebras 
naturally arise there. 

In Section 5 we show that, besides (\ref{0.7}), the factorization method can be 
applied to the following equation 
\begin{equation}
  m_{tt}=2 m*m_t+m_t*m-m*(m*m)+c_1(m_t-m*m)+c_2 m, \label{0.8}
\end{equation}
where $c_i\in \Bbb C$. In a particular case when 
$*$ is a Lie operation, (\ref{0.8}) has been linearized in [7]. 

Moreover, equations (\ref{0.7}) (see [7]) and (\ref{0.8}) can be reduced to
linear  equations with variable coefficients in the following more general case. 
Let (\ref{0.2}) be a vector space decomposition, where   
$\frak G_+$ and $\frak G_-$ are not subalgebras as before, but 
vector spaces satisfying the following conditions 
\begin{equation}
  [\frak H, \, \frak G_+] \subset \frak G_+, \qquad 
  [\frak H, \, \frak G_-] \subset \frak G_-, \label{0.9}  
\end{equation}
with 
\begin{equation}
  \frak H=[\frak G_+, \ \frak G_+]_{-}+[\frak G_-, \ \frak G_-]_{+}.
  \label{0.10}
\end{equation}
It is easy to verify that $\frak H$ is a subalgebra in $\frak G$. If 
$\frak G_+$ and $\frak G_-$ are subalgebras, then $\frak H=\{ 0\}$ 
and the conditions (\ref{0.9}) become trivial. 

As usual, the $M$-reduction is defined by a vector space $M$ such that 
\begin{equation}
  [M, \, \frak G_+] \subset M. \label{0.11}
\end{equation}

\begin{example}
Let us consider the case $\frak G=sl(n)$, \, $\frak
G_+=so(n).$ Given an expansion $n=m_1+m_2+...+m_k$, we denote by 
$\frak G_-$ a vector space of block matrices $a=\{ a_{ij}\}$
whose entries $a_{ij}$ are  $m_i \times m_j$-matrices  
such that all $a_{ii}$ are symmetric and $a_{ij}=0$ if $i>j$. 
It easy to verify that $\frak H$ coincides with the Lie algebra of 
block-diagonal skew-symmetric matrices and (\ref{0.9}) is fulfilled. One can
take  for $M$ the set of all symmetric matrices from $sl(n).$   
 
In a particular case $m_1=m_2=\cdots =m_n=1$ the vector space $\frak G_-$ 
coincides with the Lie algebra of all upper-triangular matrices, 
$\frak H=\{ 0\}$ and we deal with an ordinary decomposition of $sl(n)$ 
into a direct sum of two subalgebras.  
\end{example}

\section{$M$-reductions and operator Yang-Baxter equations}

\subsection{The case of $\Bbb Z_2$-graded Lie algebras}

Let 
\begin{equation}
  \frak G=\frak G_0\oplus \frak G_1 \label{1.1}
\end{equation} 
be a 
$\Bbb Z_2$-graded Lie algebra. In other words, the following commutation 
relations  
\begin{equation*}
  [\frak G_0, \, \frak G_0] \subset \frak G_0, \qquad
  [\frak G_0, \, \frak G_1] \subset \frak G_1, \qquad [\frak G_1, \, \frak G_1] 
  \subset \frak G_0
\end{equation*}
hold.

Let us consider an $M$-reduction (\ref{0.5}), where  $\frak G_+ =\frak G_0,$   
$M=\frak G_1$. It is clear that
\begin{equation}
  \frak G_-=\{ m-R(m) \vert m\in M\}. \label{1.2}
\end{equation}
One can show that $G_-$ is a subalgebra iff $R$ satisfies the modified 
Yang-Baxter equation 
\begin{equation}
  R\big([R(X), \ Y] - [R(Y), \ X]\big) - [R(X), \ R(Y)] - [X, \ Y]=0, 
  \qquad X,Y \in \frak G_1.\label{1.3}
\end{equation}
It is important to note that in our case $R$ is an operator defined on 
$\frak G_1$ and acting from 
$\frak G_1$ to $\frak G_0$,  whereas usually (see [4]) $R$ is required to 
be an operator on $\frak G$.

It also can be shown that $\frak G_-$ satisfies (\ref{0.9}) iff 
\begin{equation*}
  [D(X,Y),R(Z)]=R([D(X,Y),Z])
  \qquad X,Y,Z \in M,
\end{equation*}
where 
\begin{equation*}
  D(X,Y)=R\big([R(X), \ Y] - [R(Y), \ X]\big) - [R(X), \ R(Y)] - [X, \ Y], 
  \qquad X,Y \in \frak G_1. 
\end{equation*}

It is clear that for any $\frak G_+$ - invariant subspace $V \in M$ 
the corresponding $V$-reduction can be solved, 
as well as the original $M$-reduction (\ref{0.5}), by the factorization method. 
This means that if $M$ is reducible then (\ref{0.5}) admits 
further reductions. It is natural to expect that the most interesting 
reductions are related to the cases when $M$ is irreducible 
with respect to $\frak G_+$-action. 

The structure of corresponding $\Bbb Z_2$-graded Lie algebras is 
described in the following statement (see [8]). 
\begin{theorem} 
Let $\frak G$ be a 
$\Bbb Z_2$-graded Lie algebra. 
The representation of $\frak G_0$ on $\frak G_1$ is both faithful and 
irreducible if and only if $\frak G$ belongs to one of the following 
three types:
\begin{roster}
\item $\frak G$ is a simple Lie algebra; 
\item $\frak G = \frak F \oplus \frak F$, with $\frak F$ being a simple 
Lie algebra and $\frak G_0 = \{(X,X)\}, \ \  \frak G_1 = \{(X,-X)\}$; 
\item $[\frak G_1, \ \frak G_1] = 0$ and the 
action of $\frak G_0$ on $\frak G_1$ is faithful and irreducible.
\end{roster}
\end{theorem}

It follows from (\ref{1.3}) that in case (3) of Theorem 1 the operator $R$ 
satisfies the classical Yang-Baxter equation  
\begin{equation}
  R\big([R(X), \ Y] - [R(Y), \ X]\big) - [R(X), \ R(Y)]=0 \label{1.4}
\end{equation}
on $\frak G_1$. Under the additional condition that $\frak G_1=\frak G^*_0$ 
the last equation has been investigated in [2],[3]. The  
corresponding equation (\ref{0.5}) on $\frak G^*_0$ is related to a decomposition  
of the double $\frak G_0\oplus \frak G^*_0$ into a sum of two subalgebras and 
can be solved by the factorization method.  
 
In case (2) there exists the one-to-one 
correspondence between 
complementary subalgebras (\ref{1.2}) and constant solutions of the  
modified Yang-Baxter equation (\ref{1.3}) on $\frak F$. 
Equation (\ref{0.5}) on $\frak F$ was treated in [4] without a consideration of the 
double $\frak F\oplus \frak F$, but as we have seen above it 
can be solved by the standard factorization method on this double. 

For all interesting examples in case 1 we have $\frak G_1\ne \frak G^*_0.$ 
Such non-hamiltonian $\frak G_1$- reductions do not seem to be considered 
before. 

Below we present several concrete examples of equations (\ref{0.5}), related 
to different cases of Theorem 1.

\begin{example}
The Lie algebra $\frak G=sl(3)$ admits decomposition (\ref{1.1}), 
where  
\begin{equation*}
  \frak G_0 = \left\{ \left( \matrix
  a,&c,&0 \\ 
  d,&b,&0 \\ 
  0,&0,&-a-b \endmatrix \right) \right\}; \qquad \frak G_1 =
  \left\{ \left( \matrix
  0,&0,&P \\ 
  0,&0,&Q \\ 
  R,&S,&0 \endmatrix \right) \right\}.
\end{equation*}
Apparently, we are under conditions of Theorem 1, case (1). 
Let us choose the complementary subalgebra $\frak G_-$ as follows
\begin{equation*}
  \frak G_- = \left\{ \left( \matrix
  -Y+X+\alpha U,&X,&Y-X \\ 
  -Z+(2-3\alpha )U,&(1-2\alpha )U,&Z+(3\alpha -2)U \\ 
  -Y+X+U,&X,&Y-X+(\alpha -1)U \endmatrix \right) \right\},
\end{equation*}
where $\alpha$ is a (complex) parameter.

In this case, (\ref{0.5}) yields the following system of ODEs   
\begin{equation}
  \cases 
  P_t = P^2 -RP-QS \\
  Q_t = (\beta -2)RQ + \beta PQ \\
  R_t = R^2 - RP -QS \\
  S_t = (3-\beta )RS + (1-\beta )PS 
  \endcases,
  \qquad
  \beta = 3\alpha,\label{1.5}
\end{equation}
with respect to entries of the matrix 
\begin{equation*}
  U= \left\{ \left( \matrix
  0,&0,&P \\ 
  0,&0,&Q \\ 
  R,&S,&0 \endmatrix \right) \right\}.
\end{equation*}

It follows from (\ref{0.1}) that $I_1= tr \ U^2 = RP+QS$ is a first integral 
for (\ref{1.5}). Other first integrals of (\ref{0.1}) which can be obtained 
in a standard way [1] become trivial under the $M$-reduction. 
Nevertheless it is not hard to integrate (\ref{1.5}) by quadratures. 
Two additional first integrals are of the form 
\begin{equation*}
  I_2 = \frac{P-R}{QS}, \qquad I_3= Q^{1-\beta }S^{-\beta }(R^2-RP-QS).
\end{equation*}

It is interesting to note that for generic $\beta$ the system (\ref{1.5}) probably 
does not pass the Painlev\'e-Kovalevskaya test.  
\end{example}

\begin{example}
Let us consider a $\Bbb Z_2$-graded Lie algebra 
$\frak G = \frak G_0 \oplus \frak G_1$, where 
\begin{equation*}
  \frak G_0 = \left\{ \left( \matrix
  a,&c,&0,&0 \\ 
  b,&-a,&0,&0 \\ 
  0,&0,&a,&c \\
  0,&0,&b,&-a \endmatrix \right) \right\}; \qquad \frak G_1 =
  \left\{ \left( \matrix
  a,&c,&0,&0 \\ 
  b,&-a,&0,&0 \\ 
  0,&0,&-a,&-c \\
  0,&0,&-b,&a 
  \endmatrix \right) \right\}. 
\end{equation*}
It is clear that $\frak G$ is isomorphic to $sl(2) \oplus sl(2)$  
and, as a $\Bbb Z_2$-graded Lie algebra, belongs to the class (2) of 
Theorem 1. The top, corresponding (up to scaling) to 
\begin{equation*}
  \frak G_- = \left\{ \left( \matrix
  \alpha X,&Y,&0,&0 \\ 
  0,&-\alpha X,&0,&0 \\ 
  0,&0,&-X,&0 \\
  0,&0,&Z,&X \endmatrix \right) \right\}, \qquad \alpha \ne -1,
\end{equation*}
is given by 
\begin{equation}
  \cases 
  P_t = RQ \\
  Q_t = PQ \\
  R_t = \alpha RP.
  \endcases\label{1.6}
\end{equation}
The functions 
\begin{equation*}
  I_1=QR - \frac{\alpha +1}{2} P^2, \qquad I_2=R Q^{-\alpha}
\end{equation*}
are first integrals for (\ref{1.6}). 
\end{example}

\begin{example}
Let us take 
\begin{equation*}
  \left\{ \left( \matrix
  *,&*,&*,&* \\ 
  *,&*,&*,&* \\ 
  *,&*,&*,&* \\
  0,&0,&0,&0 \endmatrix \right) \right\},
\end{equation*}
for $\frak G$. It is clear that $\frak G = \frak G_0 
\oplus \frak G_1$, where 
\begin{equation*}
  \frak G_0 = \left\{ \left( \matrix
  *,&*,&*,&0 \\ 
  *,&*,&*,&0 \\ 
  *,&*,&*,&0 \\
  0,&0,&0,&0 \endmatrix \right) \right\}; \qquad \frak G_1 =
  \left\{ \left( \matrix
  0,&0,&0,&P \\ 
  0,&0,&0,&Q \\ 
  0,&0,&0,&R \\
  0,&0,&0,&0 
\endmatrix \right) \right\}. 
\end{equation*}
Obviously, the algebra $\frak G$ belongs to the class (3) of Theorem 1. 
Let $\frak G_+=\frak G_0$ and 
\begin{equation*}
  \frak G_- = \left\{ \left( \matrix
  c,&\lambda c,&a,&a \\ 
  -\lambda c,&c,&b,&b \\ 
  a,&b,&c,&c \\
  0,&0,&0,&0 \endmatrix \right) \right\}.
\end{equation*}
Since $\frak G_-$ is not a subalgebra, we have to find the vector space $H$ 
using (\ref{0.10}). A simple calculation shows that    
\begin{equation*}
  \qquad \frak H =
  \left\{ \left( \matrix
  0,&-d,&0,&0 \\ 
  d,&0,&0,&0 \\ 
  0,&0,&0,&0 \\
  0,&0,&0,&0 
  \endmatrix \right) \right\}, 
\end{equation*}
and the conditions (\ref{0.9}) are fulfilled. The corresponding equation 
(\ref{0.5}) is (up to scaling) of the form 
\begin{equation*}
  \cases 
  P_t = 2 PR + \lambda QR \\
  Q_t = 2 Q R - \lambda PR \\
  R_t = P^2+Q^2+R^2.
  \endcases
\end{equation*}
The last equation can be linearized with the help of the generalized 
factorization method [7].  
\end{example}

\subsection{One more version of the operator Yang-Baxter
equation}

In [9] (see also [10]) one more version   
\begin{equation}
  R\big([R(X), \ Y] - [R(Y), \ X]\big) - [R(X), \ R(Y)] - R^2([X, \ Y])=0 
  \label{1.7}
\end{equation}
of the operator Yang-Baxter equation on a Lie algebra $\frak F$ has been 
considered. 
In a particular case when $\frak F$ is the infinite dimensional 
Lie algebra of vector fields, (\ref{1.7}) means that the 
Fr\"olicher-Nijenhuis tensor [11] of the affinor $R$ is equal to zero. 

It turns out (see [9]) that equation 
(\ref{0.5}), where $R:\frak F \rightarrow \frak F$ satisfies (\ref{1.7}), 
can be also solved with the help of the factorization method. Namely, 
let $\frak G$ be a (finite-dimensional) factor-algebra of the polynomial 
Lie algebra $\frak F[\lambda]$ with respect to an ideal, generated by the 
minimal polynomial of the operator $R.$ For the vector space $M$ and 
the subalgebras  
$\frak G_+$ and $\frak G_-$ we take, correspondingly, 
the images of $\lambda \frak F$ and subalgebras $\frak F$ and 
\begin{equation*}
  \{ \sum_{i=1}^{m} \lambda^i (a_i-\lambda^{-1} R(a_i)) \, \vert 
  \quad a_i\in \frak F, \qquad m>0\}
\end{equation*}
under the canonical homomorphism.     
Then the $M$-reduction of (\ref{0.1}) coincides with the equation under 
consideration.

It was shown in [9], that if a Lie algebra $\frak G$ is decomposed into a 
direct sum of subalgebras $\frak G_1,...,\frak G_k$ such that the sum of 
any two subalgebras is a subalgebra, then the following operator  
$R=\lambda_1 \pi_1+...+\lambda_k \pi_k,$
where $\lambda_i\in \Bbb C$ and $\pi_i$ denotes the projector onto  
$\frak G_i,$ satisfies (\ref{1.7}). 

\begin{example}
It easy to verify that for a decomposition  
$sl(2)=\frak G_1\oplus \frak G_2\oplus \frak G_3$, where 
\begin{equation*}
  \frak G_1 = \left\{ \left( \matrix a,&0 \\ 0,&-a 
  \endmatrix \right) \right\}; 
  \qquad \frak G_2 =\left\{ \left( \matrix b,&2b \\ 0,&-b \endmatrix \right) 
  \right\},
  \qquad \frak G_3 =\left\{ \left( \matrix -c,&0 \\ 2c,&c \endmatrix \right) 
  \right\}, 
\end{equation*}
all these conditions hold. Let us define the operator $R$ as follows 
$R=\lambda_1 \pi_1+\lambda_2 \pi_2+\lambda_3 \pi_3.$
Then the corresponding equation (\ref{0.5}) for entries of the matrix   
\begin{equation*}
  m=\left( \matrix P,&Q \\ R,&-P 
  \endmatrix \right)
\end{equation*}
has the form 
\begin{equation*}
  \cases 
  P_t = (\nu-\mu) QR \\
  Q_t = 2 \mu PQ + \mu Q^2 + \nu QR \\
  R_t = -2 \nu PR - \nu R^2 - \mu QR,
  \endcases
\end{equation*}
with $\mu=\lambda_2-\lambda_1,$ \, $\nu=\lambda_3-\lambda_1$. 
A series of examples generalizing Example 5, has been presented in [9].
\end{example}

\section{Nonassociative algebras and $M$-reductions}

We associate with each $N$-dimensional algebra $\frak A$,
defined by the structural constants $C^{i}_{jk}$, a top-like system of
ODEs of the form
\begin{equation}
   u^{i}_{t}=C^{i}_{jk}u^{j}u^{k}, \qquad i,j,k = 1,\ldots,N . \label{2.1}
\end{equation}
Here and in the sequel we assume that summation from 1 to $N$ is carried
out over repeated indices. In terms of the $\frak A$-operation $*$ the
system (\ref{2.1}) takes the form
\begin{equation}
  U_t = U * U,\label{2.2}
\end{equation}
where $U(t)$ is an  $\frak A$-valued function. The system (\ref{2.1})
will be called the $\frak A$-top.

One of our purposes is to investigate relationships between properties
of $\frak A$ and the integrability of (\ref{2.1}). Recall that in this paper 
the term "integrable" means integrable by the factorization or generalized 
factorization method.

\subsection{Left-symmetric algebras}

Let us consider the left-symmetric tops. Recall that an algebra
$\frak A$ is said to be {\bf left-symmetric} (see [12]-[15]) if the
multiplication $*$ in $\frak A$ satisfies the identity
\begin{equation}
   As(X,Y,Z)=As(Y,X,Z),\label{2.3}
\end{equation}
where
\begin{equation*}
   As(X,Y,Z)=(X*Y)*Z-X*(Y*Z).
\end{equation*}

\begin{example}
Given a vector $C$, 
a vector space $\frak V$ with the operation
\begin{equation*}
   X*Y=(X,Y)C+(X,C)Y, 
\end{equation*}
where $(\cdot,\cdot)$ is the ordinary 
dot product, gives us an example of a left-symmetric algebra (see [16]).
\end{example}

\begin{example}
Let $R: \frak G \rightarrow \frak G$ be a solution of the classical operator 
Yang-Baxter equation (\ref{1.4}) on a Lie algebra $\frak G$. Then $\frak G$ 
is a left-symmetric algebra with respect to the operation (\ref{0.6}).  
\end{example}

\begin{example}
If $\frak A$ is an associative algebra and 
$R:\frak A \rightarrow \frak A$ satisfies the modified Yang-Baxter equation  
(\ref{1.3}), then the operation  
\begin{equation*}
  a*b=ab+ba+[R(a), \, b]
\end{equation*}
is a left-symmetric one. 
\end{example}

\begin{example}
The previous formula can be generalized to the case of 
arbitrary Jordan algebra $J$ with a multiplication $\circ$. Let $Der(J)$ be 
the Lie algebra of all derivations of 
$J$ and $R: J \rightarrow Der(J),$ an operator satisfying (\ref{1.3}) in the 
structural algebra $Lie(J)$ (see [17]). Then 
\begin{equation*}
  a*b=a\circ b+R(a)(b) 
\end{equation*}
is a left-symmetric operation on $J$. 
\end{example}

Let $\frak A$ be a left-symmetric algebra. It follows from (\ref{2.3}) that 
the operation $[X,Y]= X*Y-Y*X$ is a Lie bracket. Moreover, it is easy to 
verify that the vector space $\frak G = \frak A \oplus \frak A$ is a Lie 
algebra with respect to the bracket
\begin{equation}
  \big[(g_1,a_1), \ (g_2,a_2) \big] = \big( [g_1,g_2], \ g_1*a_2-g_2*a_1 \big)
  . \label{2.4}
\end{equation}
 
The Lie algebra $\frak G$ admits a $\Bbb Z_2$-gradation: $\frak G = \frak G_0
\oplus \frak G_1,$ where  $\frak G_0=\{(a,0)\}$, \, $\frak G_1= \{(0,b)\}$. 
Since $[\frak G_1, \, \frak G_1]=\{0\},$ we deal with case 3 of 
Theorem 1. 

It follows from (\ref{2.4}) that $\frak G_+=\frak G_1$ and
$\frak G_-=\{(a,-a)\}$ are subalgebras in $\frak G$. 
Equation (\ref{0.1}), corresponding to the decomposition $\frak G = \frak G_+
\oplus \frak G_-$ has the form
\begin{equation}
  V_t=U*V-V*U, \qquad U_t=V*U + U*U,\label{2.5}
\end{equation}
where $q = (V,U)$.
To obtain (\ref{2.2}) as $M$-reduction of (\ref{2.5}) we can choose $M = \frak G_1$
(i.e. put $V=0$).

\subsection{$G$-algebras and associated tops}

\resetfootnoterule

In this subsection we consider algebras with the identities 
\footnote{The identity (\ref{2.6}) means that
the operation $[X,Y]= X*Y-Y*X$ is a Lie bracket.}
\begin{gather}
   [X,Y,Z] + [Y,Z,X] + [Z,X,Y] = 0, \label{2.6} \\
   V*[X,Y,Z] = [V*X,Y,Z] + [X,V*Y,Z] + [X,Y,V*Z],\label{2.7}
\end{gather}
where
\begin{equation}
   [X,Y,Z] = As(X,Y,Z) - As(Y,X,Z).\label{2.8}
\end{equation}

We call them $G$-algebras. It is clear that left-symmetric algebras 
belong to the class of $G$-algebras.

\begin{example}
Evidently, any matrix $X$ can be decomposed 
into a sum of a skew-sym\-me\-tric matrix $X_+$ and an upper-triangular 
matrix $X_-$. Let us consider the operation (cf. (\ref{0.6})) 
$X*Y = [X_+, \ Y].$
One can verify that the set of all symmetric matrices equipped 
with this operation is a $G$-algebra. 
\end{example}
 
\begin{theorem} \ 
\begin{roster} 
\item Let $\frak G = \frak G_0\oplus \frak G_1$ be a $\Bbb Z_2$-graded 
Lie algebra. Given a splitting (\ref{0.2}),
where $\frak G_+=\frak G_0$ and $\frak G_-$ is a Lie 
subalgebra, we define an algebraic structure on $\frak G_1$ 
by the formula 
\begin{equation}
  X*Y = [X_+, \ Y],\label{2.9}
\end{equation}
where $X_+$ denotes the projection of $X$ onto $\frak G_+$ 
parallel to $\frak G_-$. Then $\frak G_1$ is a $G$-algebra with respect to 
(\ref{2.9}).
\item Any $G$-algebra can be obtained from a suitable $\Bbb Z_2$-graded 
Lie algebra by the above construction.
\end{roster}
\end{theorem}

\begin{proof}
To establish the identity (\ref{2.6}) it suffices to show 
that the operation 
\begin{equation*}
  X\times Y=[X_+, \ Y] - [Y_+, \ X], \qquad X,Y \in \frak G_1
\end{equation*}  
satisfies the Jacoby identity. In accordance with (\ref{0.2}), 
any element $X \in \frak G$ can be 
uniquely represented as $X = X_+ + X_-$, where $X_+ \in \frak G_+$, 
\ $X_- \in \frak G_-$. Let $X,Y \in \frak G_1.$ Using the formula  
\begin{equation*}
  [X_-, \ Y_-] = [X-X_+, \ Y-Y_+] = [X, \ Y] + [X_+, \ Y_+] 
  + [Y_+, \ X] - [X_+, \ Y],
\end{equation*}
and the fact that $\frak G_-$ is a Lie algebra we conclude 
that  
\begin{equation}
  [X_-, \ Y_-]_{\frak G_{1}} = Y \times X, \label{2.10}
\end{equation}
where  the  index  "$\frak G_1$"  on  the  left  hand  side  means  the 
projection onto $\frak G_1$ parallel to  
$\frak G_0$. If we apply (\ref{2.10}) to project the Jacoby identity for
$X_-, \ Y_-, \ Z_- \in \frak G_-$ onto $\frak G_1$, we make certain that 
the Jacobi identity for the operation "$\times $" holds. 

In order to prove (\ref{2.7}) we need the following relation 
\begin{equation}
  [X, \ Y, \ Z]= [[X, \ Y], \ Z], \qquad X,Y,Z \in \frak G_1, \label{2.11}
\end{equation}
where the left hand side is defined by (\ref{2.8}). It can be checked by a direct 
calculation. Rewriting (\ref{2.7}) in terms of $\frak G$-bracket 
with the help of (\ref{2.11}) we see that (\ref{2.7}) follows from the Jacobi 
identity for $\frak G$. 

It remains to prove the second part of Theorem 2. Let $\frak G_1$  
be a $G$-algebra. Put 
\begin{equation}
  \frak G = \frak G_0 \oplus \frak G_1, \label{2.12}
\end{equation}
where $\frak G_0$ is a Lie algebra generated by all operators of left 
multiplication in $\frak G_1$. Recall that the left multiplication 
operator $L_X$ is defined as follows \ 
$L_X : \frak G_1 \rightarrow X*\frak G_1$. The vector space 
$\frak G$ becomes a Lie algebra if we define 
\begin{equation}
  \big[(A,\ X), \ (B, \ Y)\big] = \big([A,B] -[L_X, \ L_Y] +L_{X*Y} - 
  L_{Y*X}, \, A(Y) -B(X) \big). \label{2.13}
\end{equation}
Obviously, the bracket (\ref{2.13}) is skew-symmetric. One can easily show that the 
identities (\ref{2.6}), (\ref{2.7}) are equivalent to the Jacobi identity for 
(\ref{2.13}). It follows from (\ref{2.13}) that the decomposition (\ref{2.12}) is a 
$\Bbb Z_2$-gradation.  To define a decomposition (\ref{0.2}) we take for 
$\frak G_-$ the set $\{(-L_X, \ X)\}$ and $\frak G_0$ for $\frak G_+$. 
As we see from (\ref{2.13}), $\frak G_-$ is a 
subalgebra in $\frak G$. For $\frak G_-$ and $\frak G_+$ thus defined,  
(\ref{2.9}) is of the form $(0,X)*(0,Y)= [(L_X,0), (0,Y)]$. 
This relation is fulfilled according to (\ref{2.13}).
\end{proof}

Note that in Example 10, $\frak G_0, \ \frak G_1, \ \frak G_-$, and 
$\frak G$ are the sets of skew-symmetric, symmetric, upper-triangular and all 
matrices, respectively. 
 
The decompositions (\ref{0.2}) from Examples 2 and 3 lead to structures of 
$G$-algebras on the corresponding vector spaces $\frak G_1.$

\section{The tops associated with $SS$-algebras}

\subsection{A relationship between $SS$-algebras and 
$\Bbb Z_2$-graded Lie algebras}

An algebra with identity 
\begin{equation}
   [V,X,Y*Z] - [V,X,Y]*Z - Y*[V,X,Z] = 0,\label{3.1}
\end{equation}
where $[X,Y,Z]$ is defined by (\ref{2.8}), 
will be called  an $SS$-algebra.  

\begin{remark}
It follows from (\ref{3.1}) that for any $SS$-algebra $\frak A$ the operator
\begin{equation*}
  K_{YZ} = [L_Y, \ L_Z] - L_{Y*Z} + L_{Z*Y} 
\end{equation*}
is a derivation of $\frak A$ for any $Y, Z$. As before, 
$L_X$ denotes the operator of left-multiplication by $X$. 
\end{remark}

Note that associative, left-symmetric, Jordan and Lie algebras are 
$SS$-algebras. 

\pagebreak[3]
\begin{theorem} \ 
\begin{roster} 
\item Let $\frak G = \frak G_0 \oplus \frak G_1$ be a 
$\Bbb Z_2$-graded Lie algebra, such that $[\frak G_1, \ \frak G_1] = 0$. 
Given  a vector space decomposition 
$\frak G = \frak G_+ \oplus \frak G_-$   
with $\frak G_+=\frak G_0$ and  
a vector space $\frak G_-$ satisfying (\ref{0.9}),   
let us equip $\frak G_1$  with an algebraic structure  
with the help of the formula (\ref{2.9}). Then the operation $*$  
satisfies the identity (\ref{3.1}).

\item Any $SS$-algebra $\frak A$ can be obtained from a suitable 
$\Bbb Z_2$-graded Lie algebra by the above construction. 
\end{roster}
\end{theorem}

\begin{proof}
We do not give a complete proof of Theorem 3. 
Its first part can be proved in the same manner as the first 
part of Theorem 2. We explain only how to construct $\frak G, \ \frak G_+, 
\ \frak G_-$  for a given $SS$-algebra. We take for $\frak G_+$ 
the Lie algebra $End \ \frak A$ of all linear endomorphisms of $\frak A$. 
The vector space $\frak G = (End \ \frak A) \oplus \frak A$ becomes a 
$\Bbb Z_2$-graded Lie algebra if we define 
\begin{equation*}
  \big[(A,\ X), \ (B, \ Y)\big] = \big([A,B], \ A(Y) -B(X) \big). 
\end{equation*}
It is not difficult to show that (\ref{3.1}) implies that 

a) the vector space $\frak H$ generated by all elements of the form 
\begin{equation*}
  \big([L_Y, \ L_Z] - L_{Y*Z} + L_{Z*Y}, \ 0\big) 
\end{equation*} 
is a Lie subalgebra in $\frak G,$ and  

b) the vector space $\frak G_- = \{(-L_X, \ X)\}$ and  
defined above subalgebras $\frak G_+$ and $\frak H$ satisfy (\ref{0.9}).  
\end{proof}

\subsection{Low-dimensional $SS$-algebras}

Straightforward computations, using the identity (\ref{3.1}), prove the following 
\begin{theorem}
Up to change of basis $e_1, e_2$, every 
non-trivial two-dimensional $SS$-algebra 
is equivalent to one of the following algebras (for a fixed value of 
non-removable parameter $\lambda$):   
\begin{align*}
\frak A_1: \qquad &e_1 * e_1 = e_1, 
           \quad e_1 * e_2 = (1 - \lambda) e_1  + \lambda e_2 ,\\
	    &e_2 * e_1 = \lambda e_1  + (1 - \lambda) e_2 , 
	    \quad e_2 * e_2 = 0; 
	\\
\frak A_2: \qquad &e_1 * e_1 = e_1, 
           \quad e_1 * e_2 = \lambda e_2,\\
	&e_2 * e_1 = 0, 
	\quad e_2 * e_2 = 0; 
	\\
\frak A_3: \qquad &e_1 * e_1 = e_1, 
           \quad e_1 * e_2 = \lambda e_2,\\
	&e_2 * e_1 = e_2, \quad e_2 * e_2 = 0;
	\\
\frak A_4: \qquad &e_1 * e_1 = 0, \quad e_1 * e_2 = \lambda e_1,\\
	&e_2 * e_1 = e_1, \quad e_2 * e_2 = e_1 + (\lambda + 1) e_2 ;
	\\
\frak A_5: \qquad &e_1 * e_1 = 0, \quad e_1 * e_2 = e_1,\\
	&e_2 * e_1 = -e_1, \quad e_2 * e_2 = 0;
	\\
\frak A_6: \qquad &e_1 * e_1 = e_1, 
	\quad e_1 * e_2 = 0,\\
	&e_2 * e_1 = 0, \quad e_2 * e_2 = e_2;
	\\
\frak A_7: \qquad &e_1 * e_1 = e_1, 
	\quad e_1 * e_2 = e_1  + 2 e_2 ,\\
	&e_2 * e_1 = 2 e_1  + e_2, \quad e_2 * e_2 = e_2;
	\\
\frak A_8: \qquad &e_1 * e_1 = 0, \quad e_1 * e_2 = e_2,\\
	&e_2 * e_1 = 0, \quad e_2 * e_2 = 0;
	\\
\end{align*}
\end{theorem}

\begin{comments}
If $\lambda = 0, 1$ then $\frak A_1$ is an associative algebra. For
$\lambda  =  1/2$ it is a Jordan algebra. 

The algebras $\frak A_2$ and $\frak A_3$ are associative at 
$\lambda = 0,1.$
 
$\frak A_4$ is a left-symmetric algebra for  $\lambda = 0$ and is an  
LT-algebra (see [17]) if  $\lambda = 1$.

$\frak A_5$ is a Lie algebra. 

$\frak A_6$ is a commutative associative algebra.

The algebras $\frak A_7$ and $\frak A_8$ are left-symmetric. 
\end{comments}

\noindent
The most non-trivial $\frak A$-top
\begin{equation*}
  {dX\over dt} = X^2  + 3 XY, \qquad        {dY\over dt} = 3 XY + Y^2
\end{equation*}
corresponds to $\frak A_7$.     

For dimension 3, we present two one-parameter families of $SS$-algebras 
with the multiplication table of the form  
\begin{align*}
  &e_1*e_2= \alpha e_3, \qquad e_2*e_1= \beta e_3, 
  \qquad e_1*e_3= \gamma e_1, \\ 
  &e_3*e_1= \delta e_1, \qquad e_2*e_3= \varepsilon e_2, 
  \qquad e_3*e_2= \mu e_2, \qquad e_3*e_3= \nu e_3. 
\end{align*}

The first is given by 
\[
  \alpha = \lambda, \qquad \beta = 2-\lambda, 
  \qquad \gamma = 0, \qquad \delta = 2, \qquad \varepsilon = 0, 
  \qquad \mu = -2, \qquad  \nu = 0. 
\]

The second is equivalent to the $SS$-algebra defined by the decomposition 
from Example~4. This corresponds to   
\[
  \alpha = 1, \qquad \beta = 1, 
  \qquad \gamma = \lambda, \qquad \delta = 2-\lambda, 
  \qquad \varepsilon = \lambda, 
  \qquad \mu = 3 \lambda -2, \qquad  \nu = \lambda. 
\]

\section{Integrable second order ODEs, related to  
$M$-reductions}
 
Let (\ref{0.2}) be a decomposition of a Lie algebra $\frak G$ such that 
the vector subspaces  $\frak G_+$ and $\frak G_-$ satisfy  
(\ref{0.9}),(\ref{0.10}) and $M$ be a vector subspace satisfying (\ref{0.11}). 

It turns out that the following initial problem 
\begin{equation*}
  X(0)=X_0, \qquad Y(0)=Y_0
\end{equation*}
for equation 
\begin{equation}
  X_t = X*X + \alpha X + \beta Y, \qquad Y_t = X*Y + \gamma X + 
  \delta Y, \qquad X,Y\in M  \label{4.1}
\end{equation}
where $\alpha, \beta, \gamma, \delta\in \Bbb C$, 
$\beta \ne 0,$ reduces to linear equations with variable coefficients 
by the generalized factorization method [7]. 
As usual, the operation * is defined by the formula (\ref{2.9}).
  
Note, that expressing $Y$ from the first equation and substituting 
into the second one can reduce (\ref{4.1}) to  
the following second order equation  
\begin{equation*}
  X_{tt} = 2 X*X_t + X_t*X - X*(X*X) + c_1 (X_t-X*X) +c_2 X,
\end{equation*}
where $c_1 = \alpha + \delta, \ c_2 = \beta \gamma- \alpha \delta$. 

At first let us solve an auxiliary system 
\begin{equation}
  U_t = \alpha U + \beta V, \qquad V_t = [U, V] + \gamma U + \delta V,
  \label{4.2}
\end{equation}
with initial data
\begin{equation*}
  U(0)=X_0, \qquad V(0)=Y_0.
\end{equation*}
Here $U,V\in \frak G.$ It is clear that this system is equivalent to 
\begin{equation}
  U_{tt} = [U, U_t] + c_1 U_t +c_2 U.\label{4.3}
\end{equation}
  
The following linearization procedure for (\ref{4.3}) was presented in [7].  
Let $Q(t)\in \frak G$ be a solution of the linear equation  
\begin{equation*}
  Q_{tt}=c_1 Q_t+c_2 Q, 
\end{equation*} 
where $Q(0)=U(0), Q_t(0)=U_t(0).$ Let us define $Y(t)$ as a solution of the 
initial problem  
\begin{equation*}
  Y_t=Y Q(t), \qquad Y(0)=E. 
\end{equation*} 
Then $U(t)=Y Q(t) Y^{-1}$ is a solution of (\ref{4.3}). 

Let us look for a solution of (\ref{4.1}) in the form 
\begin{equation*}
  X=A(t) U(t) A^{-1}(t), \qquad Y=A(t) V(t) A^{-1}(t),  
\end{equation*}
where $A(t)$ satisfies the equation  
\begin{equation}
  A_t=-\left( A U(t) A^{-1} \right)_- A, \qquad A(0)=E \label{4.4}
\end{equation}
on the Lie group $G$ of the Lie algebra $\frak G.$ Here $U(t), \ V(t)$ 
is already known solution of the system (\ref{4.2}), and "-" means the 
projection onto $\frak G_-$ parallel to $\frak G_+.$ 

If $\frak G_+$ and $\frak G_-$ are subalgebras, then (\ref{4.4}) can be reduced to  
the following factorization problem (cf. (\ref{0.4})):
\begin{equation*}
  C(t)=A^{-1}(t) B(t),
\end{equation*}
where $C_t=U(t) C$, $A(0)=B(0)=C(0)=E,$ $A(t)$ and $B(t)$ belong to the Lie 
groups of the Lie algebras $\frak G_-$ and $\frak G_+.$

In more general case, when $\frak G_-$ and $\frak G_+$ are vector 
spaces and conditions (\ref{0.9}) are fulfilled, the equation (\ref{4.4}) has been 
linearized in [7]. 

\subsection*{Acknowledgements} 
The authors thank M.A. Semenov-Tian-Shansky and 
G. Marmo for fruitful discussions. The first author (I.G) was supported by 
the Russian Fund for Basic Research (grant 99-01-00294).
The second author (V.S) was supported, in part, by RFBR grant 99-01-00294, 
INTAS project 99-1782 and the Research Programme of the Carlos III University, 
Madrid (ref. 00993). He is grateful to the Mathematics Department at this 
University and personally to Alberto Ibort Latre for their hospitality.

\label{lastpage}

\end{document}